\documentclass[11pt]{article}

\usepackage[preprint]{acl}

\usepackage{times}
\usepackage{latexsym}
\usepackage[T1]{fontenc}
\usepackage[utf8]{inputenc}
\usepackage{microtype}
\usepackage{inconsolata}
\usepackage{amsmath,amssymb,booktabs,array,tabularx,graphicx}
\usepackage{placeins,needspace,float}
\usepackage{tikz}
\usetikzlibrary{arrows.meta,positioning,calc}

\hypersetup{
    colorlinks=true,
    citecolor=blue,
    linkcolor=blue,
    urlcolor=blue,
    pdfauthor={Yangyang Qu, Michele Panariello, Massimiliano Todisco, Nicholas Evans},
    pdftitle={PAID: Phoneme-Aligned Acoustic Units for Cross-Corpus L1-Background Classification}
}

\title{
    PAID: Phoneme-Aligned Acoustic Units for Cross-Corpus L1-Background Classification
}

\author{
    Yangyang Qu \\
    EURECOM \\
    Sophia Antipolis, France
    \And
    Michele Panariello \\
    EURECOM \\
    Sophia Antipolis, France
    \AND
    Massimiliano Todisco \\
    EURECOM \\
    Sophia Antipolis, France
    \And
    Nicholas Evans \\
    EURECOM \\
    Sophia Antipolis, France
}

\begin{document}
\raggedbottom
\maketitle
\begin{abstract}
First-language (L1) background classification from second-language English speech aims to identify pronunciation-related acoustic cues that remain informative across speakers and corpora.
Global pooling is commonly used to aggregate frame-level speech representations, but averaging features across an utterance may obscure local acoustic differences in how individual phonemes are pronounced.
To model these pronunciation-related cues before utterance-level aggregation, we introduce Phoneme-Aligned Input Design (PAID), which constructs acoustic units from frozen Whisper representations using transcript-assisted forced alignment.
PAID applies a shared nonlinear mapping to each acoustic unit before utterance-level aggregation and incorporates auxiliary phoneme-label embeddings.
We construct a six-L1 dataset with speaker-disjoint partitions and evaluate cross-corpus performance on the Speech Accent Archive (SAA).
PAID achieves $89.72\%$ Macro-F1 on the source Test set and $80.14\%$ on SAA.
Matched comparisons show that phoneme-aligned acoustic-unit modeling improves SAA Macro-F1 by $13.44$ percentage points over global pooling.
To address uneven classification performance across L1 backgrounds, we additionally evaluate class-sensitive training, which improves observed weakest-class performance.
\end{abstract}

\section{Introduction}
\label{sec:introduction}
First-language (L1) background classification aims to identify a speaker's L1 from speech produced in a second language (L2).
For L2 English, differences in the acoustic realization of individual speech sounds can provide cues related to a speaker's L1 background \citep{shao2022,yang2023}.
The resulting challenge is how to aggregate pronunciation-related acoustic cues into an utterance-level L1 prediction.

Pretrained speech encoders provide frame-level representations that can be used for accent-related classification \citep{zuluaga2023commonaccent}.
A common approach uses global mean pooling to obtain a fixed-dimensional utterance representation.
In a pooling-first classifier, the learned nonlinear mapping receives an utterance-wide average rather than separate segment-level acoustic inputs.
We therefore investigate whether preserving segment-level acoustic inputs until after a learned transformation improves L1-background classification.

Phonetic structure has previously been used in L1-background classification through phone-level statistics and sub-phone acoustic modeling \citep{abad2016,qian2017}.
Phoneme alignment provides estimated boundaries for grouping contextual speech representations according to phoneme occurrences.
Because the same phoneme category can have different acoustic realizations \citep{choi2025}, we retain continuous acoustic features within each aligned span rather than replacing them with discrete phoneme labels.
This provides a linguistically informed alternative to global frame pooling for utterance-level classification.

We introduce \textbf{Phoneme-Aligned Input Design (PAID)} to model these acoustic inputs before utterance-level aggregation.
Using transcript-assisted alignment from the Montreal Forced Aligner (MFA) \citep{mcauliffe2017} at training and inference, PAID averages frame representations from a frozen Whisper-small encoder \citep{radford2023robust} within each retained non-silence phoneme span.
Each resulting vector is a phoneme-aligned acoustic unit corresponding to one phoneme occurrence.
A shared learned nonlinear mapping transforms each unit before utterance-level aggregation.
PAID additionally incorporates phoneme-label embeddings to summarize label composition, while the acoustic-only configuration retains the aligned acoustic units but omits these embeddings.

To evaluate PAID across speakers and corpora, we construct a six-L1 classification dataset by curating metadata from L2-ARCTIC \citep{zhao2018} and ESLTTS \citep{wang2024}, harmonizing L1 labels, and defining speaker-disjoint Train, Development, and Test partitions.
We further evaluate cross-corpus performance on the Speech Accent Archive (SAA) \citep{weinberger2015speech} under the same six-class label space.
PAID trained with class-weighted cross-entropy (Weighted CE) achieves $89.72\%$ Macro-F1 on source Test and $80.14\%$ on SAA, exceeding the highest-scoring evaluated non-PAID baseline on SAA by $11.78$ percentage points.

To examine the modeling choices behind these results, we conduct matched comparisons using the same frozen Whisper encoder and training settings.
Under ordinary cross-entropy (CE), MFA spans (acoustic-only) improve SAA Macro-F1 over global pooling by $13.44$ percentage points.
Under the same acoustic-only architecture, MFA-derived spans outperform duration-matched artificial spans by $6.31$ points.
These comparisons support the overall acoustic-unit design and the choice of MFA-derived segmentation over the evaluated artificial alternative.

We additionally examine class-sensitive training to assess performance differences across L1 backgrounds.
Mean--max CE increases mean SAA Worst-F1 from $63.16\%$ to $71.00\%$ relative to ordinary CE, using the same PAID architecture and input construction.

Our contributions are threefold:
\begin{itemize}
    \item We introduce PAID, a transcript-assisted L1-background classifier built on phoneme-aligned acoustic units.

    \item We construct a six-L1 benchmark with harmonized labels and speaker-disjoint partitions, complemented by cross-corpus evaluation under the same label space.

    \item We use matched comparisons to distinguish the effects of acoustic-unit modeling, MFA-derived segmentation, and phoneme-label embeddings, and analyze the impact of class-sensitive training.
\end{itemize}

\section{Related Work}
\label{sec:related-work}

\paragraph{L1-background classification and pretrained speech representations.}
Native-language identification from second-language speech has been studied using phone log-likelihood ratios, sub-phone acoustic modeling, and neural spectrogram-based classifiers \citep{abad2016,qian2017,graham2021}.
Other studies have investigated phonetic and prosodic information in accent recognition \citep{shao2022,yang2023}.
More recent work has explored pretrained speech representations for accent and dialect classification, including CommonAccent \citep{zuluaga2023commonaccent}, Voxlect \citep{feng2026voxlect}, and Vox-Profile \citep{feng2025voxprofile}.
Our work instead focuses on constructing and aggregating phoneme-aligned acoustic units from frozen contextual speech representations for L1-background classification.

\paragraph{Phoneme-aware acoustic modeling and aggregation.}
Pretrained speech encoders have been shown to capture phonetic information \citep{choi2024}.
Phoneme-level supervision has been explored for adapting speech representations to spoken language modeling \citep{poli2024improving}.
Phoneme-dependent acoustic distributions have also been investigated for modeling pronunciation variation \citep{choi2025}.
For representation aggregation, Deep Sets establishes a general framework based on shared transformations and permutation-invariant pooling \citep{zaheer2017deepsets}.
Unlike phoneme-supervised encoder adaptation, PAID uses transcript-derived alignment to structure frozen contextual acoustic representations without introducing a phoneme-classification objective.

\paragraph{Class-sensitive optimization.}
Class weighting and balanced sampling are commonly used to address class imbalance, while worst-group optimization emphasizes groups with higher prediction losses \citep{sagawa2020}.
Related min--max formulations combine average losses with worst-group terms in subgroup-fairness learning \citep{pereira2025m2fgb}.
Building on this general principle, we evaluate Mean--max cross-entropy (CE), which combines utterance-average CE with the highest class-mean CE in each minibatch, treating L1 labels as groups.
We examine its effect on overall and weakest-class classification performance.

\section{Method}
\label{sec:task-model}

\begin{figure*}[t]
\centering
\includegraphics[width=\textwidth]{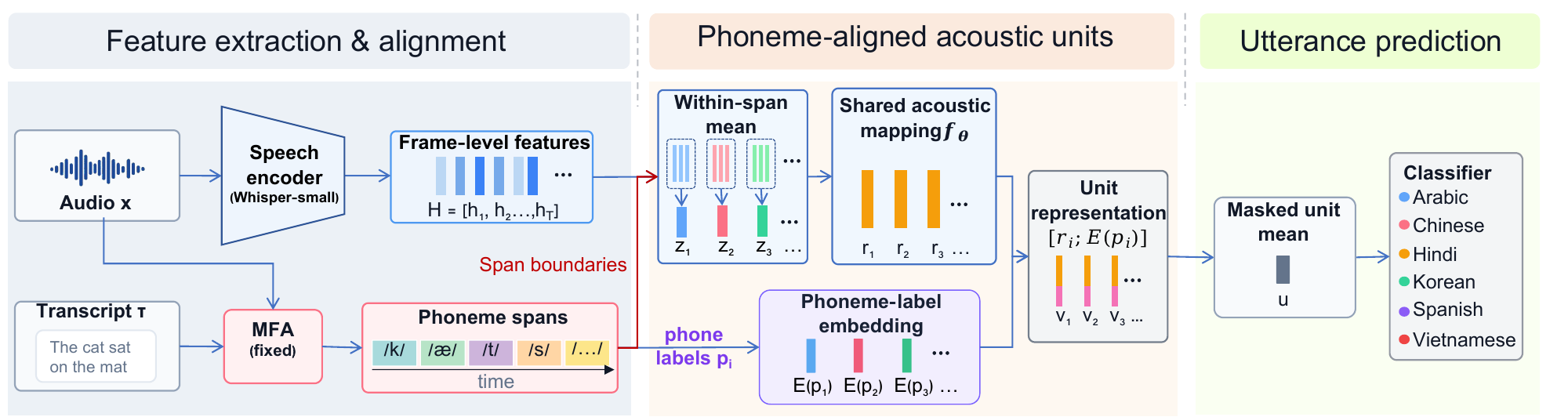}
\caption{\textbf{PAID architecture.}
A frozen Whisper-small encoder extracts contextual frame representations, while transcript-assisted forced alignment provides non-silence phoneme spans.
PAID averages the encoder features within each span to construct an acoustic unit, applies a shared nonlinear mapping to each unit, and aggregates the transformed units for utterance-level classification.
The phoneme-label branch, included in full PAID and omitted in acoustic-only controls, summarizes phoneme-label composition without conditioning the acoustic mapping on phoneme identity.}
\label{fig:model}
\end{figure*}

Figure~\ref{fig:model} illustrates PAID, which constructs phoneme-aligned acoustic units from frozen speech representations and transforms them before utterance-level aggregation.

\subsection{Task and Phoneme Alignment}
\label{sec:l1-task}

Given an L2 English waveform $x$ and its transcript $\tau$, the task is to predict the speaker's L1-background label $y\in\{1,\ldots,C\}$, where $C$ is the number of L1 classes.
A frozen Whisper-small encoder \citep{radford2023robust} extracts contextual frame representations
$
H=(\mathbf{h}_1,\ldots,\mathbf{h}_T),
\mathbf{h}_t\in\mathbb{R}^{768},
$
at approximately 20-ms intervals, where $T$ denotes the number of valid encoder frames corresponding to the processed audio, excluding padding.

We use the Montreal Forced Aligner (MFA) \citep{mcauliffe2017} to obtain phoneme-level alignments from the waveform and transcript.
The retained non-silence phoneme spans are represented as
$
A=\{(s_i,e_i,p_i)\}_{i=1}^{K},
\label{eq:phoneme-spans}
$
where $s_i$ and $e_i$ are the start and end times of span $i$, $p_i$ is its assigned phoneme label, and $K$ is the number of retained spans used for acoustic-unit construction.
The label $p_i$ is assigned through transcript-assisted phoneme alignment, whereas $y$ denotes the utterance-level L1 background to be predicted.

MFA alignment is performed using transcripts during both training and inference, and the resulting boundaries remain fixed during classifier training.
The boundaries determine which encoder frames are grouped into each acoustic unit.
The assigned phoneme labels are used as classifier inputs only when the optional phoneme-label branch is enabled.

\subsection{Phoneme-Aligned Acoustic Units}
\label{sec:aligned-acoustic-representation}

We convert the MFA timestamps into Whisper encoder-frame indices and restrict the resulting spans to valid, non-padding frames.
For each retained span $i$, let $I_i\subseteq\{1,\ldots,T\}$ denote its assigned frame indices, with $|I_i|>0$.
We construct a phoneme-aligned acoustic unit by averaging the contextual frame representations within the span:
\begin{equation}
\mathbf{z}_i
=
\frac{1}{|I_i|}
\sum_{t\in I_i}\mathbf{h}_t.
\label{eq:acoustic-unit}
\end{equation}
Each $\mathbf{z}_i\in\mathbb{R}^{768}$ summarizes the contextual acoustic features associated with one aligned phoneme occurrence.
Although the pooling region is restricted to a phoneme span, its input features remain contextual representations produced by the frozen encoder.

We apply a shared nonlinear mapping $f_\theta$ to each acoustic unit:
$
\mathbf{r}_i
=
f_\theta(\mathbf{z}_i)
\in\mathbb{R}^{192}.
\label{eq:phoneme-units}
$
The mapping consists of a linear projection, layer normalization, ReLU activation, and dropout.
The same mapping is applied to every acoustic unit within an utterance, allowing the classifier to transform individual span representations before combining them into an utterance-level representation.

\paragraph{Comparison with global pooling.}
We compare global pooling with phoneme-aligned acoustic-unit aggregation:
\begin{equation}
\begin{aligned}
\mathbf{u}_{\mathrm{global}}
&=f_\theta\!\left(
\frac{1}{T}\sum_{t=1}^{T}\mathbf{h}_t
\right),\\
\mathbf{u}_{\mathrm{aligned}}
&=\frac{1}{K}\sum_{i=1}^{K}f_\theta(\mathbf{z}_i).
\end{aligned}
\label{eq:aggregation-order}
\end{equation}

Whisper-Global applies the learned mapping to the utterance-wide frame average, whereas MFA spans (acoustic-only) transform each aligned acoustic unit before averaging.
Both configurations use the same mapping architecture with separately learned parameters.
This comparison evaluates the overall acoustic-unit design, including the pooling regions, aggregation weights, and order of mapping and averaging, rather than isolating any single operation.

\subsection{Utterance Prediction and Phoneme-Label Composition}
\label{sec:explicit-phone-identity}
We consider PAID and an acoustic-only configuration to examine the contribution of explicit phoneme-label information.
Both use the same MFA-derived acoustic-unit construction and acoustic mapping architecture.
The acoustic-only configuration uses only the transformed acoustic representations $\mathbf r_i$, whereas PAID additionally incorporates a trainable 64-dimensional embedding $E(p_i)$ of each aligned phoneme label:
\begin{equation}
\mathbf{v}_i=
\begin{cases}
\mathbf{r}_i, & \text{acoustic-only},\\
[\mathbf{r}_i;E(p_i)], & \text{PAID}.
\end{cases}
\label{eq:phone-token}
\end{equation}
Here, $[\cdot;\cdot]$ denotes vector concatenation.
Averaging the phoneme-label embeddings summarizes their relative occurrence frequencies within the utterance, providing complementary information to the acoustic units.

For minibatch processing, let $m_i\in\{0,1\}$ indicate whether position $i$ corresponds to a valid acoustic unit, and let $M=\sum_i m_i>0$ denote the number of valid units in the utterance.
We compute the utterance representation through masked mean pooling and obtain the classification logits using dropout followed by a linear layer:
\begin{equation}
\mathbf{u}
=
\frac{1}{M}\sum_i m_i\mathbf{v}_i,
\qquad
\mathbf{o}
=
W\,d(\mathbf{u})+\mathbf{b},
\label{eq:paid-classifier}
\end{equation}
where $d$ denotes dropout, $W\in\mathbb{R}^{C\times D}$, and $\mathbf{b}\in\mathbb{R}^{C}$.
The classifier input dimension is $D=192$ for the acoustic-only configuration and $D=256$ for PAID.
The predicted L1 class is obtained by applying softmax to the logits and selecting the class with the highest probability.

For the acoustic-only configuration, the utterance representation is $\mathbf{u}_{\mathrm{aligned}}$.
For PAID, the utterance representation concatenates the pooled acoustic representation with the mean phoneme-label embedding:
\begin{equation}
\mathbf{u}
=
\left[
\mathbf{u}_{\mathrm{aligned}};
\frac{1}{K}\sum_{i=1}^{K}E(p_i)
\right].
\label{eq:phone-composition}
\end{equation}
Because the phoneme-label embeddings are incorporated after the shared acoustic mapping, they do not condition $f_\theta$ on phoneme identity.
Instead, they provide an auxiliary summary of phoneme-label composition, which contributes additively to the classification logits when dropout is disabled (Appendix~\ref{app:phone-composition}).

\subsection{Training Objectives}
\label{sec:paid-training}

The Whisper encoder and MFA-derived alignments remain fixed throughout classifier training.
We optimize the acoustic mapping $f_\theta$, the phoneme-label embeddings when enabled, and the final classification layer.
For an utterance $j$ with L1 label $y_j$, the cross-entropy (CE) loss is
$\ell_j=-\log P(y_j\mid x_j,\tau_j)$.

\paragraph{Class-weighted cross-entropy.}
The default PAID configuration uses class-weighted CE (Weighted CE) to account for differences in the number of training examples across L1 classes.
For class $c$, we define $w_c=N/(Cn_c)$, where $N$ is the total number of training examples and $n_c$ is the number belonging to class $c$.
For a minibatch $B$, the weighted objective is
\begin{equation}
\mathcal{L}_{\mathrm{wCE}}
=
\frac{\sum_{j\in B}w_{y_j}\ell_j}
{\sum_{j\in B}w_{y_j}}.
\label{eq:weighted-ce}
\end{equation}
Ordinary CE assigns equal weight to individual training examples and is used for the matched comparisons of global pooling, artificial segmentation, and phoneme-aligned acoustic units unless otherwise specified.

\paragraph{Class-sensitive Mean--max CE.}
Class weighting adjusts example contributions according to class frequency, which does not necessarily reflect the current classification difficulty of each class.
We additionally evaluate a Mean--max CE objective motivated by worst-group optimization \citep{sagawa2020} and mixed mean--worst-group objectives \citep{pereira2025m2fgb}.
Treating the L1 classes as groups, this objective combines utterance-average CE with the highest class-mean CE among the classes present in each minibatch.

Let $B_c=\{j\in B:y_j=c\}$ denote the examples belonging to class $c$ in minibatch $B$.
We define
\begin{equation}
\mathcal{L}_{\alpha}
=
(1-\alpha)\frac{1}{|B|}\sum_{j\in B}\ell_j
+
\alpha\max_{c:|B_c|>0}
\frac{1}{|B_c|}\sum_{j\in B_c}\ell_j.
\label{eq:objective-mixture}
\end{equation}
Here, $\alpha\in[0,1]$ controls the contribution of the highest class-mean loss.
Setting $\alpha=0$ recovers ordinary CE, while $\alpha=1$ retains only the highest class-mean CE among the classes present in the minibatch.
Intermediate values balance the two terms.

Mean--max CE emphasizes classes with higher minibatch-average prediction losses but does not directly optimize class-wise F1.
We compare the training objectives using the same frozen encoder features, phoneme-span construction, and downstream architecture, with trainable parameters optimized separately for each configuration.

\section{Evaluation Setup}
\label{sec:controls-protocols}

\subsection{Dataset Construction and Evaluation Sets}
\label{sec:main-protocol}

We construct a six-L1 classification dataset from L2-ARCTIC \citep{zhao2018} and ESLTTS \citep{wang2024} by curating recordings and speaker metadata, harmonizing L1 labels, and defining speaker-disjoint Train, Development (Dev), and Test partitions.
The dataset covers Arabic, Chinese, Hindi, Korean, Spanish, and Vietnamese, with both source corpora represented in every L1 class and partition.
Table~\ref{tab:dataset-summary} reports the recordings retained after alignment and acoustic-feature preprocessing.
For cross-corpus evaluation, we use 420 recordings from distinct speakers in the Speech Accent Archive (SAA) \citep{weinberger2015speech}, covering the same six L1 backgrounds.
The source-trained classifiers are evaluated directly on SAA; alignment-based configurations use the available transcripts for both source and SAA evaluation.
Detailed corpus coverage and split information are provided in Appendices~\ref{app:dataset-coverage} and~\ref{app:provenance}.

\begin{table}[t]
\centering
\small
\begin{tabular*}{\columnwidth}{@{\extracolsep{\fill}}lrr@{}}
\toprule
Partition & Speakers & Utterances \\
\midrule
Train & 57 & 31,372 \\
Dev & 16 & 12,187 \\
Test & 18 & 11,097 \\
\midrule
SAA (external) & 420 & 420 \\
\bottomrule
\end{tabular*}
\caption{\textbf{Evaluation data.}
Train, Dev, and Test are speaker-disjoint partitions of the six-L1 dataset constructed in this study.
Counts reflect recordings retained after preprocessing; SAA is used for cross-corpus evaluation.}
\label{tab:dataset-summary}
\end{table}

\begin{table*}[t]
\centering
\small
\setlength{\tabcolsep}{4pt}
\caption{\textbf{Matched Whisper-based representation controls.}
All configurations use the same frozen Whisper encoder and matched training protocol.
Acoustic-only configurations exclude explicit phoneme-label embeddings.}
\label{tab:representation-conditions}
\begin{tabularx}{\textwidth}{@{}p{0.22\textwidth}p{0.36\textwidth}p{0.12\textwidth}X@{}}
\toprule
Configuration & Acoustic input and aggregation & Label embeddings & Role in the comparison \\
\midrule
Whisper-Global
& Mean over non-padding encoder frames, followed by $f_\theta$
& No
& Global-pooling baseline \\
Artificial spans (acoustic-only)
& Mean within reconstructed spans, followed by $f_\theta$ and mean over spans
& No
& Duration-matched segmentation control \\
MFA spans (acoustic-only)
& Mean within MFA spans, followed by $f_\theta$ and mean over spans
& No
& Phoneme-aligned acoustic-only reference \\
PAID
& Mean within MFA spans, followed by $f_\theta$ and mean over spans
& Yes
& Additional phoneme-label information \\
\bottomrule
\end{tabularx}
\end{table*}

\subsection{Baselines and Representation Controls}
\label{sec:representation-contract}

\paragraph{Pretrained-representation baselines.}
We compare PAID with six-class classifiers built on frozen representations from Whisper-small, WavLM Base+ \citep{chen2022wavlm}, ECAPA-TDNN \citep{desplanques2020ecapa}, CommonAccent \citep{zuluaga2023commonaccent}, Voxlect-small \citep{feng2026voxlect}, and Vox-Profile \citep{feng2025voxprofile}.
For each pretrained model, we extract frozen speech representations and train a separate six-class classifier on the dataset constructed in this study.
Whisper-Utt and WavLM-Utt use multilayer perceptron (MLP) classifiers on utterance-level mean-pooled features, while their logistic regression (LR) variants use the same features. Whisper-Global is a separately trained, matched Whisper control used to evaluate the acoustic-unit design under the shared training protocol. WavLM-FrameAttn instead uses frame-level attention.
CommonAccent and ECAPA-Spk use logistic-regression classifiers, while Voxlect-small and Vox-Profile use MLP classifiers.
PAID-WavLM uses frozen WavLM features in place of Whisper features while retaining PAID's acoustic-unit construction and classification architecture.
We use these systems for cross-model comparison and the matched Whisper-based controls to assess the downstream acoustic-unit design.
Checkpoint and implementation details are provided in Appendix~\ref{app:baseline-implementation}.

\paragraph{Matched representation controls.}
We compare four configurations using the same frozen Whisper-small encoder, training examples, optimization schedule, and five matched seeds, as summarized in Table~\ref{tab:representation-conditions}.
For the artificial-span control, we rearrange the original MFA span durations within each utterance while preserving the span count, duration collection, and ordered inter-span gap durations.
The reconstructed boundaries are used to recompute acoustic units rather than reorder existing representations (Appendix~\ref{app:paired-representation}).
All four configurations are evaluated under ordinary CE, with additional Weighted CE comparisons for Whisper-Global, MFA spans (acoustic-only), and PAID.

\subsection{Training and Evaluation}
\label{sec:training-details}

PAID and the matched Whisper-based controls are trained with AdamW \citep{loshchilov2019decoupled} for ten epochs using five matched seeds, with results reported from the final epoch.
Voxlect-small and Vox-Profile follow the same classifier-training schedule; the other pretrained baselines use their recorded Dev-selected results, and logistic-regression baselines have one reported run.
Throughout the paper, \emph{PAID} denotes the architecture with phoneme-label embeddings; its default training objective is Weighted CE, while alternative objectives are indicated explicitly.
For class-sensitive training, we compare ordinary CE, Weighted CE, Mean--max CE ($\alpha=0.25$), and class-, corpus-, and corpus--class-balanced sampling using the same PAID architecture.
Optimization hyperparameters and additional training-objective details are provided in Appendices~\ref{app:implementation} and~\ref{app:objective-results}.

\paragraph{Evaluation metrics.}
\label{sec:metrics}
We use Macro-F1, the unweighted mean of class-wise F1 scores, as our primary metric.
We additionally report accuracy (Acc.) and balanced accuracy (BAcc.), with BAcc. defined as the average recall across the $C$ L1 classes:
\begin{equation}
\mathrm{BAcc}
=\frac{1}{C}\sum_{c=1}^{C}
\frac{\mathrm{TP}_c}{\mathrm{TP}_c+\mathrm{FN}_c},
\label{eq:balanced-accuracy}
\end{equation}
where $\mathrm{TP}_c$ and $\mathrm{FN}_c$ denote true positives and false negatives for class $c$.
We report Worst-F1 as the minimum class F1 within each run.
For five-seed experiments, each metric is computed separately for each seed and reported as the mean $\pm$ sample standard deviation (SD).
The class-wise F1 range is $G_{\mathrm{class}}=\max_c\overline{\mathrm{F1}}_c-\min_c\overline{\mathrm{F1}}_c$, where $\overline{\mathrm{F1}}_c$ is the mean F1 of class $c$ across seeds.
\section{Results and Analysis}
\label{sec:results}

\subsection{Overall and Cross-Corpus Performance}
\label{sec:overall-results}

\begin{table*}[t]
\centering
\definecolor{paidBestBlue}{HTML}{0072B2}
\definecolor{paidSecondOrange}{HTML}{C45100}
\newcommand{\paidBest}[1]{\textcolor{paidBestBlue}{\textbf{#1}}}
\newcommand{\paidSecond}[1]{\textcolor{paidSecondOrange}{#1}}
\caption{\textbf{L1-background classification with frozen speech representations.}
Scores (\%) on the speaker-disjoint source Test set and cross-corpus SAA.
Results are five-run means unless marked with $^\dagger$ (single run).
PAID (Weighted CE) is the default configuration; Mean--max CE uses $\alpha=0.25$.
Aligned configurations require transcripts and forced alignment.
Blue bold and orange indicate the highest and second-highest displayed means in each column, respectively.}
\label{tab:overall-results}
\footnotesize
\setlength{\tabcolsep}{1.8pt}
\renewcommand{\arraystretch}{1.02}
\begin{tabular*}{\textwidth}{@{\extracolsep{\fill}}lcccccccc@{}}
\toprule
& \multicolumn{4}{c}{\textbf{Test}}
& \multicolumn{4}{c}{\textbf{SAA}} \\
\cmidrule(lr){2-5}\cmidrule(lr){6-9}
Model & Acc. & Macro-F1 & BAcc. & Worst-F1
& Acc. & Macro-F1 & BAcc. & Worst-F1 \\
\midrule

\multicolumn{9}{@{}l}{\textit{General-purpose acoustic representations}} \\
Whisper-Utt
& 86.79 & 87.89 & 88.02 & 77.48
& 62.19 & 56.95 & 66.94 & 22.81 \\
Whisper-Utt-LR$^{\dagger}$
& 86.92 & 87.97 & 88.02 & 79.07
& 63.57 & 56.60 & 68.29 & 23.93 \\
WavLM-Utt
& 82.45 & 83.22 & 83.88 & 72.49
& 40.00 & 40.64 & 51.39 & 18.78 \\
WavLM-Utt-LR$^{\dagger}$
& 83.36 & 84.46 & 84.88 & 73.28
& 38.10 & 38.23 & 48.52 & 10.34 \\
WavLM-FrameAttn
& 80.53 & 80.94 & 81.83 & 70.24
& 39.29 & 38.05 & 48.76 & 17.38 \\
Whisper-Global (matched)
& 86.68 & 87.79 & 87.95 & 76.86
& 65.29 & 62.47 & 65.56 & 29.12 \\

\midrule
\multicolumn{9}{@{}l}{\textit{Speaker-pretrained representations}} \\
ECAPA-Spk$^{\dagger}$ \citep{desplanques2020ecapa}
& 70.89 & 70.10 & 72.55 & 53.51
& 40.48 & 31.09 & 34.33 & 18.05 \\

\midrule
\multicolumn{9}{@{}l}{\textit{Accent-pretrained representations}} \\
CommonAccent-ECAPA$^{\dagger}$\citep{zuluaga2023commonaccent}
& 70.86 & 71.90 & 73.49 & 55.57
& 42.14 & 35.89 & 38.85 & 15.38 \\
CommonAccent-XLSR$^{\dagger}$ \citep{zuluaga2023commonaccent}
& 73.49 & 74.77 & 75.25 & 63.93
& 21.67 & 21.21 & 32.21 & 13.70 \\
Voxlect-small \citep{feng2026voxlect}
& 79.56 & 81.40 & 81.78 & 54.62
& 61.90 & 55.24 & 57.57 & 42.85 \\
Vox-Profile \citep{feng2025voxprofile}
& 81.93 & 83.87 & 83.82 & 60.78
& 74.43 & 68.36 & 71.76 & 35.60 \\

\midrule
\multicolumn{9}{@{}l}{\textit{PAID configurations}} \\
PAID-WavLM
& 84.95 & 86.03 & 86.40 & 74.16
& 47.33 & 46.44 & 56.50 & 21.25 \\
PAID (Weighted CE)
& \paidBest{88.77} & \paidBest{89.72} & \paidBest{89.60} & \paidBest{82.56}
& \paidBest{80.81} & \paidBest{80.14} & \paidBest{79.59} & \paidSecond{69.04} \\
PAID (Mean--max CE)
& \paidSecond{87.98} & \paidSecond{89.01} & \paidSecond{88.93} & \paidSecond{80.77}
& \paidSecond{80.14} & \paidSecond{79.90} & \paidSecond{78.42} & \paidBest{71.00} \\
\bottomrule
\end{tabular*}
\end{table*}

Table~\ref{tab:overall-results} compares PAID with classifiers built on general-purpose, speaker-pretrained, and accent-pretrained speech representations.
PAID (Weighted CE) achieves $89.72\pm0.47\%$ Macro-F1 on the speaker-disjoint source Test set and $80.14\pm1.84\%$ on SAA.
On SAA, it exceeds Vox-Profile + MLP, the highest-scoring evaluated non-PAID baseline in Macro-F1 ($68.36\pm2.76\%$), by $11.78$ percentage points.
On the source Test, PAID exceeds the highest reported non-PAID Macro-F1, obtained by the single-run Whisper-Utt-LR ($87.97\%$), by $1.75$ points.
PAID (Weighted CE) also achieves the highest displayed mean accuracy and balanced accuracy on both evaluation sets, while PAID (Mean--max CE) obtains the highest mean SAA Worst-F1 ($71.00\%$).
The matched Whisper-based comparisons in Section~\ref{sec:representation-analysis} examine the contribution of PAID's acoustic-unit design.

\paragraph{Performance across L1 backgrounds.}
Table~\ref{tab:saa-classwise} compares class-wise SAA F1 for PAID trained with ordinary CE and the two accent-pretrained baselines, Voxlect-small + MLP and Vox-Profile + MLP, whose downstream classifiers also use ordinary CE.
PAID has the highest displayed mean F1 for five of the six L1 classes, while Vox-Profile + MLP scores higher for Arabic.
Relative to Vox-Profile, the largest gains are observed for Korean ($+27.56$ points), Hindi ($+12.85$), and Chinese ($+11.23$).
The range across the six class-wise mean F1 scores is $25.57$ points for PAID, compared with $30.88$ for Voxlect-small + MLP and $46.75$ for Vox-Profile + MLP.
Per-class variability is reported in Appendix~\ref{app:saa-class-variability}.

\begin{table}[t]
\centering
\footnotesize
\setlength{\tabcolsep}{4pt}
\caption{\textbf{Class-wise SAA F1 for PAID (CE) and accent-pretrained baselines.}
Values are five-seed mean F1 scores (\%).
$N$ denotes the number of SAA recordings per L1 class.
The F1 range is the difference between the highest and lowest class-wise mean F1 scores.}
\label{tab:saa-classwise}
\begin{tabular*}{\linewidth}{@{\extracolsep{\fill}}lrrrr@{}}
\toprule
L1 & $N$ & Voxlect & Vox-Profile & PAID (CE) \\
\midrule
Arabic & 101 & 44.53 & \textbf{79.83} & 76.66 \\
Chinese & 65 & 49.27 & 64.58 & \textbf{75.81} \\
Hindi & 18 & 63.94 & 75.88 & \textbf{88.73} \\
Korean & 52 & 47.29 & 35.60 & \textbf{63.16} \\
Spanish & 162 & 75.41 & 82.35 & \textbf{83.06} \\
Vietnamese & 22 & 50.99 & 71.93 & \textbf{80.79} \\
\midrule
F1 range (pp) & -- & 30.88 & 46.75 & \textbf{25.57} \\
\bottomrule
\end{tabular*}
\end{table}

\paragraph{Effect of combining source corpora.}
Training PAID with ordinary CE on our six-L1 dataset achieves $89.33\%$/$78.04\%$ Test/SAA Macro-F1, compared with $76.88\%$/$68.78\%$ using L2-ARCTIC alone and $70.25\%$/$67.83\%$ using ESLTTS alone.
To assess whether these gains result solely from more training data, we evaluate combined-source subsets matched in size to each individual corpus.
These subsets still outperform single-source training on both Test and SAA, while repeatedly sampling either corpus to match the number of examples processed per epoch does not close the gap.
These results suggest benefits from combining the two corpora beyond the evaluated effects of training-set size and sample exposure (Appendix~\ref{app:source-results}).

\subsection{What Drives PAID's Improvement?}
\label{sec:representation-analysis}

Table~\ref{tab:component-controls} reports the matched representation controls defined in Section~\ref{sec:representation-contract}.

\begin{table}[t]
\centering
\footnotesize
\setlength{\tabcolsep}{3pt}
\caption{\textbf{Matched Whisper-based representation comparisons.}
Test and SAA Macro-F1 (\%), as five-seed mean $\pm$ sample SD.
All configurations use the same frozen Whisper-small encoder and matched training schedule.
Artificial spans use duration-matched reconstructed boundaries; acoustic-only conditions omit phoneme-label embeddings.}
\label{tab:component-controls}
\begin{tabular*}{\linewidth}{@{\extracolsep{\fill}}lrr@{}}
\toprule
Representation & Test & SAA \\
\midrule
\multicolumn{3}{@{}l}{\emph{Ordinary CE}} \\
Whisper-Global & $87.79\pm0.85$ & $62.47\pm9.48$ \\
Artificial spans (acoustic-only) & $87.98\pm0.73$ & $69.59\pm3.71$ \\
MFA spans (acoustic-only) & $88.77\pm0.58$ & $75.91\pm3.78$ \\
PAID & $89.33\pm0.39$ & $78.04\pm1.20$ \\
\addlinespace[2pt]
\multicolumn{3}{@{}l}{\emph{Weighted CE}} \\
Whisper-Global & $88.07\pm0.89$ & $64.43\pm10.39$ \\
MFA spans (acoustic-only) & $89.27\pm0.49$ & $78.66\pm2.33$ \\
PAID & $89.72\pm0.47$ & $80.14\pm1.84$ \\
\bottomrule
\end{tabular*}
\end{table}

\paragraph{Phoneme-aligned acoustic-unit modeling.}
Compared with Whisper-Global, MFA spans (acoustic-only) improve SAA Macro-F1 by $13.44\pm6.39$ paired points under ordinary CE and $14.23\pm8.54$ under Weighted CE, with positive gains across all five seeds in both settings.
The improvements are more pronounced on SAA than on source Test, supporting the overall phoneme-aligned acoustic-unit design in the evaluated cross-corpus setting (Appendix~\ref{app:paired-representation}).

\paragraph{MFA-derived segmentation.}
Under the same acoustic-only architecture and ordinary CE, MFA-derived spans outperform duration-matched artificial spans by $6.31\pm0.26$ paired points on SAA and $0.80\pm0.25$ on source Test, with positive differences across all five seeds.
This supports MFA-derived segmentation over the evaluated duration-matched alternative.

\paragraph{Phoneme-label composition.}
Adding phoneme-label embeddings yields smaller, more variable paired SAA Macro-F1 gains ($2.13\pm3.91$ points under ordinary CE and $1.48\pm3.92$ under Weighted CE).
However, mean SAA Worst-F1 increases from $49.88\%$ to $63.16\%$ under ordinary CE (Appendix~\ref{app:representation-objective}).
Together, these results show that most of the measured SAA Macro-F1 gain over global pooling is already present without explicit phoneme-label embeddings, which provide complementary, metric-dependent benefits.

\paragraph{Symbolic-input controls.}
MFA spans (acoustic-only, CE) achieve $75.91\pm3.78\%$ SAA Macro-F1, compared with $9.54\pm2.08\%$ for the evaluated phoneme-label-only classifier and at most $6.58\%$ for the word-, character-, and phoneme-based term frequency--inverse document frequency (TF--IDF) classifiers.
These symbolic-only inputs do not reproduce the acoustic model's performance on the shared-passage SAA evaluation (Appendix~\ref{app:symbolic}).

\paragraph{Performance by source corpus.}
Under ordinary CE, PAID improves Macro-F1 over Whisper-Global on L2-ARCTIC ($85.55\%$ vs.\ $80.89\%$), but not on ESLTTS ($79.80\%$ vs.\ $80.45\%$; Appendix~\ref{app:source-conditional}).

\subsection{Class-Sensitive Training and Class-wise Performance}
\label{sec:objective-analysis}

Table~\ref{tab:training-objectives} compares training objectives and sampling rules with the PAID architecture and input construction held fixed.

\begin{table}[t]
\centering
\footnotesize
\setlength{\tabcolsep}{4pt}
\caption{\textbf{Training-objective and sampling comparisons on SAA.}
Five-seed mean $\pm$ sample SD (\%).
All settings use the same PAID architecture, input construction, and training schedule; model parameters are learned separately.
Mean--max CE uses $\alpha=0.25$.}
\label{tab:training-objectives}
\begin{tabular*}{\linewidth}{@{\extracolsep{\fill}}lrr@{}}
\toprule
Training rule & Macro-F1 & Worst-F1 \\
\midrule
Ordinary CE & $78.04\pm1.20$ & $63.16\pm5.73$ \\
Class-balanced sampling & $78.94\pm3.23$ & $65.62\pm5.67$ \\
Corpus-balanced sampling & $78.01\pm2.30$ & $61.70\pm6.17$ \\
Corpus$\times$class-balanced & $79.48\pm3.08$ & $67.19\pm4.08$ \\
Weighted CE & $\mathbf{80.14\pm1.84}$ & $69.04\pm3.29$ \\
Mean--max CE & $79.90\pm3.04$ & $\mathbf{71.00\pm2.97}$ \\
\bottomrule
\end{tabular*}
\end{table}

Weighted CE achieves the highest mean SAA Macro-F1 ($80.14\%$), while Mean--max CE ($\alpha=0.25$) achieves the highest mean SAA Worst-F1 ($71.00\%$) among the evaluated training rules.
Relative to ordinary CE, Mean--max CE changes Macro-F1 from $78.04\%$ to $79.90\%$ and mean Worst-F1 from $63.16\%$ to $71.00\%$.
Relative to Weighted CE, the paired Worst-F1 difference is $+1.96\pm5.67$ points.
These comparisons illustrate how the evaluated training objectives affect average and weakest-class performance differently (Appendix~\ref{app:objective-results}).

\paragraph{Class-wise effects.}
On SAA, Mean--max CE improves Korean and Chinese F1 by $10.91$ and $6.07$ points, respectively, while Arabic and Hindi decrease by $2.90$ and $3.12$ points.
The range across the six seed-averaged class F1 scores decreases from $25.57$ to $11.85$ points on SAA, while remaining similar on source Test ($16.39$ versus $16.69$ points).
These results show a change in the balance of class-wise performance rather than uniform improvements across L1 classes.
Per-class results and objective-weight sensitivity are detailed in Appendix~\ref{app:objective-results}.

\section{Conclusion}
\label{sec:conclusion}

We introduced PAID, a transcript-assisted L1-background classifier that constructs phoneme-aligned acoustic units from frozen speech representations.
By transforming each acoustic unit before utterance-level aggregation, PAID allows the classification head to process span-level acoustic inputs rather than only a global frame average.
Matched Whisper-based comparisons show a benefit from the overall aligned acoustic-unit design, while the artificial-span control supports the use of MFA-derived segmentation beyond matched duration statistics.
Complementary training-objective comparisons show that class-sensitive optimization can improve observed weakest-class performance while changing the balance of results across L1 classes.
Together, these findings support phoneme-aligned acoustic-unit modeling for the evaluated L1-background classification setting.

\section*{Limitations}
PAID requires transcripts and phoneme-level forced alignment during both training and inference, and its sensitivity to transcription or alignment errors has not been evaluated.
Our experiments cover six L1 backgrounds in L2 English, with cross-corpus evaluation conducted on the development-visible Speech Accent Archive (SAA).
The findings therefore remain specific to the evaluated languages, speakers, and recording conditions.
Comparisons with pretrained baselines involve different upstream training conditions and should be interpreted as system-level comparisons.

\section*{Ethical Considerations}
L1-background classification may reveal information about speakers' linguistic backgrounds and could be misused for discriminatory profiling.
The six L1 categories represent dataset-defined groups and should not be interpreted as comprehensive descriptions of speakers' linguistic identities.

\bibliography{references}
\clearpage
\appendix

% ============================================================
% Appendix A
% ============================================================
\appendix

% ============================================================
% Appendix A
% ============================================================
% Insert immediately after \bibliography{references}.
% Keep the appendix switch in one location; remove the old switch before insertion.
% Include once, immediately after the main paper's references.
% The main file must not contain another \appendix command.
\clearpage
\appendix

\section{Additional Representation Analyses}
\label{app:additional-representation}

\subsection{Paired Representation Comparisons and Artificial Spans}
\label{app:paired-representation}

For each of the five matched training seeds $s$, we calculate the difference in Macro-F1 between two representation configurations as $\Delta_s=\mathrm{MacroF1}_{A,s}-\mathrm{MacroF1}_{B,s}$.
Table~\ref{tab:paired-representation} reports the mean and sample standard deviation (SD) of $\Delta_s$ and the number of seeds with positive differences.
Paired differences are calculated before rounding the reported scores.

\begin{table}[!htbp]
\centering
\small
\setlength{\tabcolsep}{3pt}
\caption{\textbf{Paired representation comparisons on SAA.}
Macro-F1 differences in percentage points (pp), reported as five-seed mean $\pm$ sample SD.
Improved seeds indicates the number of matched runs in which the first configuration scores higher.
The span-based comparisons are acoustic-only and exclude phoneme-label embeddings.}
\label{tab:paired-representation}
\begin{tabularx}{\linewidth}{@{}>{\raggedright\arraybackslash}Xrr@{}}
\toprule
Comparison & $\Delta$ Macro-F1 & Improved seeds \\
\midrule
MFA spans (acoustic-only) $-$ Whisper-Global (CE) & $13.44\pm6.39$ & 5/5 \\
MFA spans $-$ artificial spans (acoustic-only, CE) & $6.31\pm0.26$ & 5/5 \\
MFA spans (acoustic-only) $-$ Whisper-Global (Weighted CE) & $14.23\pm8.54$ & 5/5 \\
\bottomrule
\end{tabularx}
\end{table}

Under ordinary CE, MFA-aligned acoustic-only units also outperform artificial spans on the source Test set by $0.80\pm0.25$ percentage points in matched-seed comparisons, with positive differences for all five seeds.
The MFA--artificial comparison holds the acoustic-only classifier architecture fixed, whereas the Whisper-Global comparison evaluates the overall acoustic-unit construction and aggregation.

\paragraph{Artificial-span construction.}
Let $(s_i,e_i)$ denote the boundaries of the $K$ retained MFA spans within one utterance, with span durations $d_i=e_i-s_i$ and inter-span gaps $g_i=s_{i+1}-e_i$.
The artificial-span control rearranges the collection of durations using a deterministic permutation $\pi$, while retaining the gaps in their original order.
The reconstructed boundaries are
\begin{equation}
\begin{aligned}
\tilde{s}_1 &= s_1, \qquad \tilde{e}_i = \tilde{s}_i+d_{\pi(i)},\\
\tilde{s}_{i+1} &= \tilde{e}_i+g_i,\qquad i=1,\ldots,K-1.
\end{aligned}
\label{eq:artificial-span-construction}
\end{equation}
This preserves the span count, the collection of span durations, the ordered gap durations, and the first onset and last offset of the retained spans.
We then convert the reconstructed boundaries to Whisper encoder-frame indices and recompute the acoustic-unit inputs from the corresponding frame features.
The control therefore changes the acoustic segmentation rather than rearranging previously extracted unit vectors.
The same time-to-frame conversion procedure is applied to MFA and artificial spans, although identical duration collections need not produce identical frame counts.
% TODO-VERIFY: Recover the exact deterministic permutation algorithm and
% generator-to-run mapping from the experiment artifacts before final submission.

\subsection{Contribution of Phoneme-Label Embeddings}
\label{app:representation-objective}

Table~\ref{tab:app-representation-objective} compares MFA spans (acoustic-only) with PAID under the same acoustic-unit construction.
The latter additionally uses phoneme-label embeddings.
All reported multi-run results use five training seeds, and Mean--max CE uses $\alpha=0.25$.

\begin{table}[!htbp]
\centering
\footnotesize
\setlength{\tabcolsep}{2pt}
\caption{\textbf{Effect of phoneme-label embeddings.}
Test and SAA scores (\%), reported as five-seed mean $\pm$ sample SD.
Mean--max CE uses $\alpha=0.25$.}
\label{tab:app-representation-objective}
\begin{tabularx}{\linewidth}{@{}>{\raggedright\arraybackslash}X>{\raggedright\arraybackslash}p{0.24\linewidth}rr@{}}
\toprule
Model & Objective & Macro-F1 & Worst-F1 \\
\midrule
\multicolumn{4}{@{}l}{\emph{Test}} \\
MFA spans (acoustic-only) & CE & $88.77\pm0.58$ & $79.19\pm1.55$ \\
MFA spans (acoustic-only) & Mean--max CE & $88.84\pm0.84$ & $80.14\pm2.59$ \\
PAID & CE & $89.33\pm0.39$ & $80.79\pm1.15$ \\
PAID & Mean--max CE & $89.01\pm0.40$ & $80.77\pm1.18$ \\
\midrule
\multicolumn{4}{@{}l}{\emph{SAA}} \\
MFA spans (acoustic-only) & CE & $75.91\pm3.78$ & $49.88\pm10.68$ \\
MFA spans (acoustic-only) & Mean--max CE & $79.08\pm1.87$ & $66.79\pm5.93$ \\
PAID & CE & $78.04\pm1.20$ & $63.16\pm5.73$ \\
PAID & Mean--max CE & $79.90\pm3.04$ & $71.00\pm2.97$ \\
\bottomrule
\end{tabularx}
\end{table}

Under ordinary CE, adding phoneme-label embeddings yields a paired SAA Macro-F1 gain of $2.13\pm3.91$ percentage points, positive in four of five seeds.
The corresponding mean SAA Worst-F1 scores are $49.88\%$ for the MFA-aligned acoustic-only configuration and $63.16\%$ for PAID.
These averages do not report the variability of the paired Worst-F1 differences.
Under Weighted CE, the paired Macro-F1 gain is $1.48\pm3.92$ points, positive in three of five seeds.
The paired Macro-F1 gains from phoneme-label embeddings are smaller and more variable than those from comparing MFA-aligned acoustic-only modeling with global pooling.

\subsection{Transcript and Phoneme-Label Controls}
\label{app:symbolic}

We evaluate symbolic-input classifiers using word and character sequences from transcripts and phoneme-label sequences obtained through MFA alignment, without Whisper encoder features.
The word, character, and phoneme $n$-gram controls use term frequency--inverse document frequency (TF--IDF) features with logistic-regression classifiers and Development-selected regularization.
Each TF--IDF configuration has one reported run with seed 1337.
The phoneme-label-only model uses learned 64-dimensional embeddings followed by masked mean pooling, dropout, and a linear six-class classifier; it is trained under ordinary CE with five seeds.
Phoneme labels obtained through MFA require both speech and transcripts, unlike the word- and character-based inputs.

\begin{table}[t]
\centering
\footnotesize
\setlength{\tabcolsep}{2pt}
\caption{\textbf{Transcript and phoneme-label controls with an acoustic reference.}
Macro-F1 (\%) on Development (Dev), Test, and SAA.
$n$ is the number of reported runs; $\pm$ denotes sample SD across five training seeds.
Dashes indicate scores not reported in the available experiment records.
The acoustic-unit reference uses ordinary CE.}
\label{tab:app-symbolic}
\begin{tabularx}{\linewidth}{@{}>{\raggedright\arraybackslash}Xrrrr@{}}
\toprule
Input / model & $n$ & Dev & Test & SAA \\
\midrule
\multicolumn{5}{@{}l}{\emph{Without Whisper encoder features}} \\
Word TF--IDF & 1 & 11.24 & 16.25 & 1.37 \\
Character TF--IDF & 1 & 13.46 & 19.81 & 1.37 \\
Phoneme n-gram TF--IDF & 1 & 27.52 & 26.95 & 6.58 \\
Phoneme-label-only & 5 & -- & -- & $9.54\pm2.08$ \\
\midrule
\multicolumn{5}{@{}l}{\emph{Acoustic reference}} \\
MFA spans (acoustic-only, CE) & 5 & -- & $88.77\pm0.58$ & $75.91\pm3.78$ \\
\bottomrule
\end{tabularx}
\end{table}

On SAA, the MFA-aligned acoustic-only model achieves $75.91\pm3.78\%$ Macro-F1, compared with $9.54\pm2.08\%$ for the phoneme-label-only model and at most $6.58\%$ for the evaluated TF--IDF controls.
Because SAA uses a shared reading passage, both word and character TF--IDF classifiers predict Hindi for all 420 recordings, of which 18 are labeled Hindi.
The phoneme $n$-gram classifier predicts 247 recordings as Hindi, 167 as Spanish, and six as Chinese.
These results characterize the tested symbolic inputs under the present evaluation protocol.

\subsection{Representation Visualization}
\label{app:representation-visualization}

\begin{figure}[!htbp]
\centering
\includegraphics[width=\columnwidth]{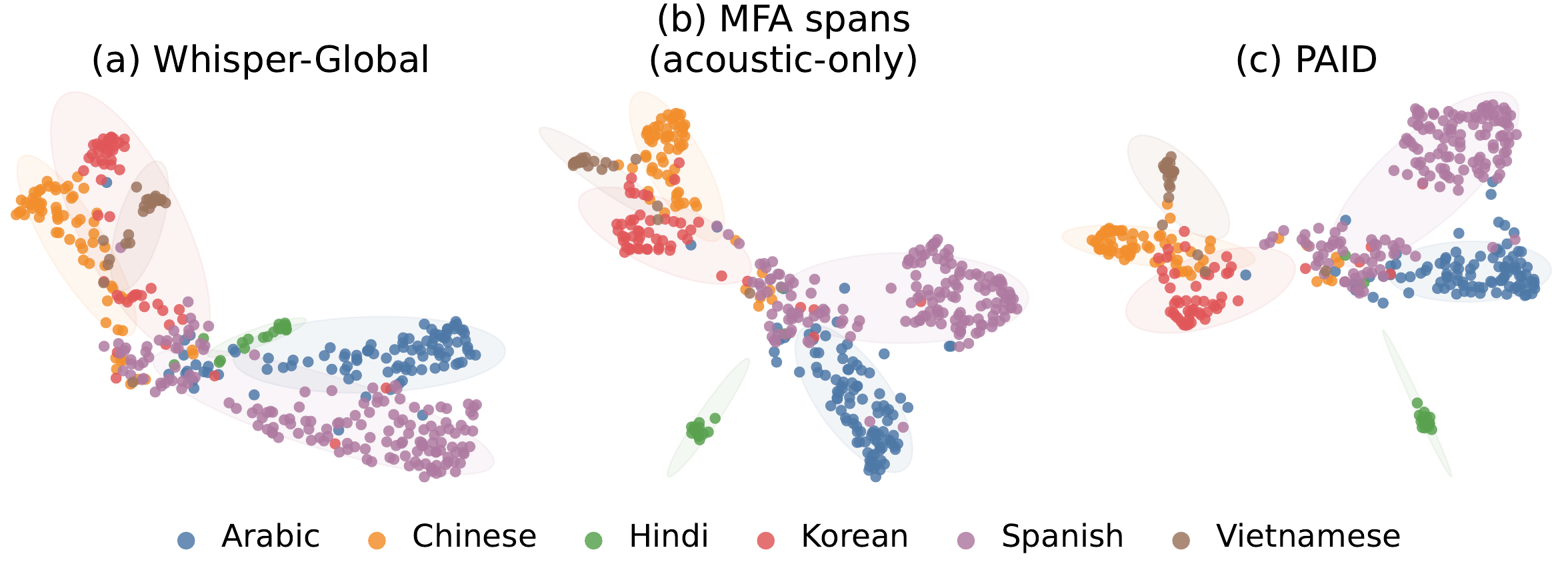}
\caption{\textbf{Utterance representations on SAA.}
Uniform Manifold Approximation and Projection (UMAP) \citep{mcinnes2018umap} of (a) Whisper-Global, (b) MFA spans (acoustic-only), and (c) PAID.
Colors denote L1 backgrounds, and ellipses summarize within-class covariance in the two-dimensional projections.}
\label{fig:representation-umap}
\end{figure}

Figure~\ref{fig:representation-umap} provides a qualitative view of utterance-level representations on SAA.

The UMAP projections use cosine distance, 40 neighbors, minimum distance 0.15, repulsion strength 1.0, spread 1.0, and projection seed 1337.
Ellipses follow the principal axes of the class covariance estimated in each two-dimensional projection.
The visualizations are descriptive and are not used as quantitative evidence of class separation.
% TODO-VERIFY: Specify the model checkpoint(s) and representation extraction
% points for each UMAP panel if these records are available.

\section{Additional System Results}
\label{app:additional-baselines}

\subsection{Other Pretrained Baselines and PAID-WavLM}

Table~\ref{tab:app-baseline-statistics} reports Macro-F1 and the number of runs for additional pretrained-representation classifiers and PAID-WavLM.
Logistic-regression configurations have one reported run; other configurations in this table have five runs.

\begin{table}[!htbp]
\centering
\footnotesize
\setlength{\tabcolsep}{2pt}
\caption{\textbf{Additional system-level Macro-F1 results.}
$n$ denotes the number of reported runs.
Multi-run values are mean $\pm$ sample SD (\%).}
\label{tab:app-baseline-statistics}
\begin{tabularx}{\linewidth}{
@{}>{\raggedright\arraybackslash}Xrrr@{}}
\toprule
Model & $n$ & Test & SAA \\
\midrule
Whisper-Utt + MLP
& 5 & $87.89\pm0.54$ & $56.95\pm4.20$ \\
Whisper-Utt + LR
& 1 & 87.97 & 56.60 \\
WavLM-Utt + MLP
& 5 & $83.22\pm0.55$ & $40.64\pm1.70$ \\
WavLM-Utt + LR
& 1 & 84.46 & 38.23 \\
WavLM-FrameAttn
& 5 & $80.94\pm0.55$ & $38.05\pm0.98$ \\
PAID-WavLM
& 5 & $86.03\pm0.67$ & $46.44\pm2.50$ \\
CommonAccent-XLSR + LR
& 1 & 74.77 & 21.21 \\
CommonAccent-ECAPA + LR
& 1 & 71.90 & 35.89 \\
ECAPA-Spk + LR
& 1 & 70.10 & 31.09 \\
\bottomrule
\end{tabularx}
\end{table}

\subsection{Accent-Pretrained Representations}
\label{app:vox-baselines}

Table~\ref{tab:app-vox-summary} reports Macro-F1 and Worst-F1 for Voxlect-small and Vox-Profile on Test and SAA, including across-seed variability.

\begin{table}[!htbp]
\centering
\small
\setlength{\tabcolsep}{3pt}
\caption{\textbf{Accent-pretrained representation results.}
Five-seed mean $\pm$ sample SD (\%).}
\label{tab:app-vox-summary}
\begin{tabular*}{\linewidth}{@{\extracolsep{\fill}}lrr@{}}
\toprule
Metric & Test & SAA \\
\midrule
\multicolumn{3}{@{}l}{\emph{Voxlect-small + MLP}} \\
Macro-F1
& $81.40\pm0.67$
& $55.24\pm1.81$ \\
Worst-F1
& $54.62\pm3.10$
& $42.85\pm3.83$ \\
\midrule
\multicolumn{3}{@{}l}{\emph{Vox-Profile + MLP}} \\
Macro-F1
& $83.87\pm0.67$
& $68.36\pm2.76$ \\
Worst-F1
& $60.78\pm3.67$
& $35.60\pm5.15$ \\
\bottomrule
\end{tabular*}
\end{table}

\subsection{Class-Wise SAA Results}
\label{app:saa-class-variability}

Table~\ref{tab:app-vox-classes} extends the main-text class-wise SAA comparison with across-seed SD.
The PAID column uses ordinary CE, matching Table~\ref{tab:saa-classwise}.

\begin{table}[!htbp]
\centering
\footnotesize
\setlength{\tabcolsep}{2pt}
\caption{\textbf{Class-wise SAA F1 with run variability.}
$N$ denotes the number of SAA utterances in each L1 class.
Values are five-seed mean $\pm$ sample SD (\%).}
\label{tab:app-vox-classes}
\begin{tabularx}{\linewidth}{@{}l r >{\centering\arraybackslash}X >{\centering\arraybackslash}X >{\centering\arraybackslash}X@{}}
\toprule
L1 & $N$ & \shortstack{Voxlect\\+ MLP} & \shortstack{Vox-Profile\\+ MLP} & \shortstack{PAID\\(CE)} \\
\midrule
Arabic & 101 & $44.53\pm4.13$ & $79.83\pm1.97$ & $76.66\pm4.58$ \\
Chinese & 65 & $49.27\pm0.74$ & $64.58\pm5.07$ & $75.81\pm3.17$ \\
Hindi & 18 & $63.94\pm5.73$ & $75.88\pm4.53$ & $88.73\pm4.88$ \\
Korean & 52 & $47.29\pm4.37$ & $35.60\pm5.15$ & $63.16\pm5.73$ \\
Spanish & 162 & $75.41\pm1.34$ & $82.35\pm1.57$ & $83.06\pm0.59$ \\
Vietnamese & 22 & $50.99\pm1.96$ & $71.93\pm8.04$ & $80.79\pm6.13$ \\
\bottomrule
\end{tabularx}
\end{table}

\section{Training-Source Analyses}
\label{app:source-controls}
\label{app:source-results}

\subsection{Single-Source and Matched-Size Training}

After preprocessing, the source training set contains 13,292 L2-ARCTIC examples and 18,080 ESLTTS examples, for 31,372 examples in the curated six-L1 training set.
Size-matched conditions subsample the curated training set to match the number of distinct examples available from either source.
Exposure-matched conditions instead sample 31,372 examples with replacement from a single source per epoch.
All conditions use the PAID architecture with ordinary CE and five training seeds.

\begin{table}[!htbp]
\centering
\footnotesize
\setlength{\tabcolsep}{2pt}
\caption{\textbf{Training-source controls.}
PAID with ordinary CE; Macro-F1 (\%), five-seed mean $\pm$ SD.
``Unified'' denotes training on the six-L1 dataset constructed from both source corpora.}
\label{tab:source-training-main}
\begin{tabularx}{\linewidth}{
@{}>{\raggedright\arraybackslash}Xrr@{}}
\toprule
Training condition & Test & SAA \\
\midrule
L2-ARCTIC only
& $76.88\pm0.18$
& $68.78\pm2.16$ \\
ESLTTS only
& $70.25\pm0.59$
& $67.83\pm4.60$ \\
Unified
& $89.33\pm0.39$
& $78.04\pm1.20$ \\
\midrule
Unified (L2-size matched)
& $87.08\pm0.70$
& $78.33\pm2.08$ \\
Unified (ESLTTS-size matched)
& $88.10\pm0.65$
& $78.92\pm1.43$ \\
L2-ARCTIC (exposure matched)
& $75.55\pm0.27$
& $67.82\pm1.85$ \\
ESLTTS (exposure matched)
& $71.12\pm0.69$
& $68.45\pm3.30$ \\
\bottomrule
\end{tabularx}
\end{table}

All evaluated training conditions using both source corpora outperform the single-source conditions on Test and SAA.
The complete curated training set yields the highest Test Macro-F1 among these conditions, while the two size-matched curated subsets yield slightly higher mean SAA Macro-F1 than the complete set.
These comparisons indicate that the observed benefit of using both sources is not reproduced by the evaluated single-source exposure-matching controls.

\subsection{Test Performance by Source Corpus}
\label{app:source-conditional}

Table~\ref{tab:source-conditional} reports the Test performance within each source corpus, rather than pooling recordings across sources.

\begin{table}[!htbp]
\centering
\footnotesize
\setlength{\tabcolsep}{2pt}
\caption{\textbf{Test performance within each source.}
Five-seed mean $\pm$ sample SD (\%).}
\label{tab:source-conditional}
\begin{tabularx}{\linewidth}{
@{}>{\raggedright\arraybackslash}Xrr@{}}
\toprule
Model / objective & Macro-F1 & Worst-F1 \\
\midrule
\multicolumn{3}{@{}l}{\emph{L2-ARCTIC}} \\
Whisper-Global (CE)
& $80.89\pm1.70$
& $39.96\pm6.16$ \\
MFA spans (acoustic-only, CE)
& $84.53\pm1.00$
& $52.28\pm4.52$ \\
PAID (CE)
& $85.55\pm0.80$
& $56.81\pm3.51$ \\
PAID (Weighted CE)
& $\mathbf{86.54}\pm0.89$
& $\mathbf{61.78}\pm3.63$ \\
\midrule
\multicolumn{3}{@{}l}{\emph{ESLTTS}} \\
Whisper-Global (CE)
& $\mathbf{80.45}\pm0.95$
& $22.17\pm4.08$ \\
MFA spans (acoustic-only, CE)
& $79.31\pm0.42$
& $19.52\pm1.98$ \\
PAID (CE)
& $79.80\pm0.34$
& $21.01\pm1.21$ \\
PAID (Weighted CE)
& $80.21\pm0.61$
& $\mathbf{22.95}\pm2.31$ \\
\bottomrule
\end{tabularx}
\end{table}

Under ordinary CE, PAID improves Macro-F1 over Whisper-Global on L2-ARCTIC ($85.55\%$ vs.\ $80.89\%$), but is slightly lower on ESLTTS ($79.80\%$ vs.\ $80.45\%$).
This indicates that the pooled Test improvement does not hold uniformly across the two source corpora.

\subsection{Source--L1 Breakdown}

Table~\ref{tab:app-source-class} gives the class-wise results and recording counts within each Test source for the default PAID (Weighted CE) configuration.

\begin{table}[!htbp]
\centering
\small
\setlength{\tabcolsep}{2pt}
\caption{\textbf{Default PAID by Test source and L1.}
F1 is five-seed mean $\pm$ sample SD (\%); $N$ denotes utterance support.}
\label{tab:app-source-class}
\begin{tabular*}{\linewidth}{
@{\extracolsep{\fill}}lrr@{}}
\toprule
L1 & $N$ & F1 \\
\midrule
\multicolumn{3}{@{}l}{\emph{L2-ARCTIC}} \\
Arabic & 1,130 & $93.44\pm0.37$ \\
Chinese & 1,132 & $77.00\pm1.13$ \\
Hindi & 1,131 & $98.77\pm0.12$ \\
Korean & 1,131 & $61.78\pm3.63$ \\
Spanish & 1,132 & $92.03\pm0.52$ \\
Vietnamese & 1,132 & $96.22\pm0.32$ \\
\midrule
\multicolumn{3}{@{}l}{\emph{ESLTTS}} \\
Arabic & 821 & $97.96\pm0.46$ \\
Chinese & 1,257 & $91.13\pm0.39$ \\
Hindi & 165 & $87.73\pm0.71$ \\
Korean & 1,283 & $95.87\pm0.21$ \\
Spanish & 632 & $85.59\pm0.82$ \\
Vietnamese & 151 & $22.95\pm2.31$ \\
\bottomrule
\end{tabular*}
\end{table}

The corresponding speaker counts are given in Table~\ref{tab:app-dataset-composition}.
Several source--L1 Test subsets contain only one held-out speaker, so their class-wise scores describe the fixed test speaker population.

\section{Training-Objective Analyses}
\label{app:objective-results}

\subsection{Training-Rule Details}
\label{app:training-objective-definition}

The Mean--max CE objective is defined in Equation~\ref{eq:objective-mixture}.
Class-balanced sampling rebalances the six L1 labels, corpus-balanced sampling rebalances the two source corpora, and corpus$\times$class-balanced sampling rebalances their intersections.
Each sampler draws examples with replacement and trains PAID using ordinary CE.

\subsection{Paired Objective Comparisons}

For each matched seed, we compute the Mean--max CE ($\alpha=0.25$) minus Weighted CE difference using the PAID architecture.
On SAA, the paired differences are $-0.24\pm1.34$ percentage points in Macro-F1 and $+1.96\pm5.67$ points in Worst-F1.
The respective differences are positive in two and four of the five seeds.

\subsection{Sensitivity to Mean--max Objective Weight}
\label{app:alpha-sensitivity}

\begin{figure}[!htbp]
\centering
\includegraphics[width=\columnwidth]{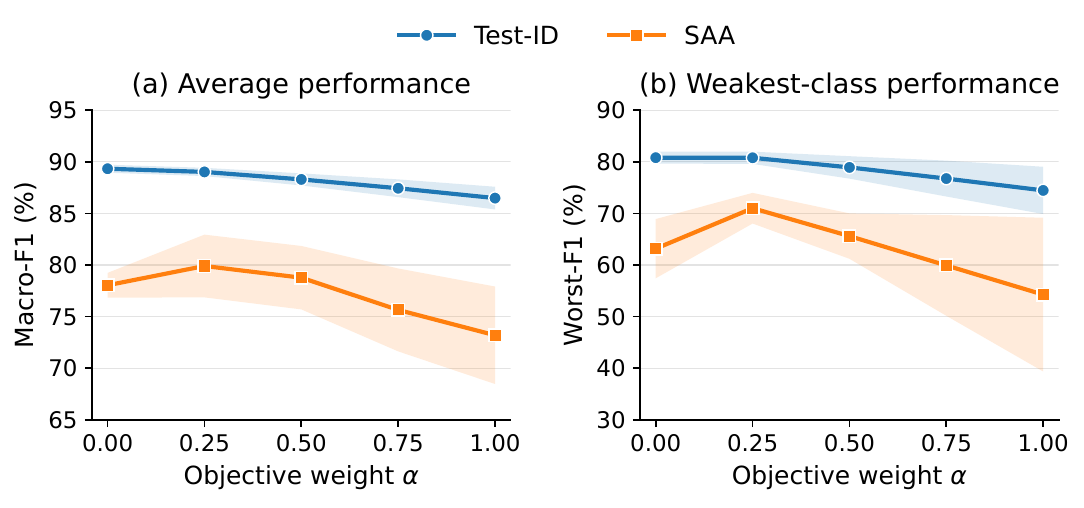}
\caption{\textbf{Sensitivity to the Mean--max CE weight $\alpha$.}
(a) Macro-F1 and (b) Worst-F1 on the source Test set and SAA across five training seeds.
Lines show mean scores, and shaded regions indicate one standard deviation across seeds.
}
\label{fig:alpha-sweep}
\end{figure}

We evaluate $\alpha\in\{0,.25,.50,.75,1\}$ while holding the PAID architecture, input construction, data, training seeds, and reporting epoch fixed.
Figure~\ref{fig:alpha-sweep} shows Test and SAA metrics for this sweep.
Among these settings, $\alpha=0.25$ yields the highest reported SAA Macro-F1 and Worst-F1, whereas $\alpha=0$ yields the highest Test Macro-F1.
The default Weighted CE and reported $\alpha=0.25$ Mean--max CE configurations were selected after examining SAA results.
% NOTE: The original source also reports the highest Development Macro-F1 at
% alpha=0; retain this claim only if its numerical source is cited elsewhere.

\subsection{Class-Wise Effects of Training Objectives}
\label{app:objective-per-class}

\begin{figure}[!htbp]
\centering
\includegraphics[width=\columnwidth]{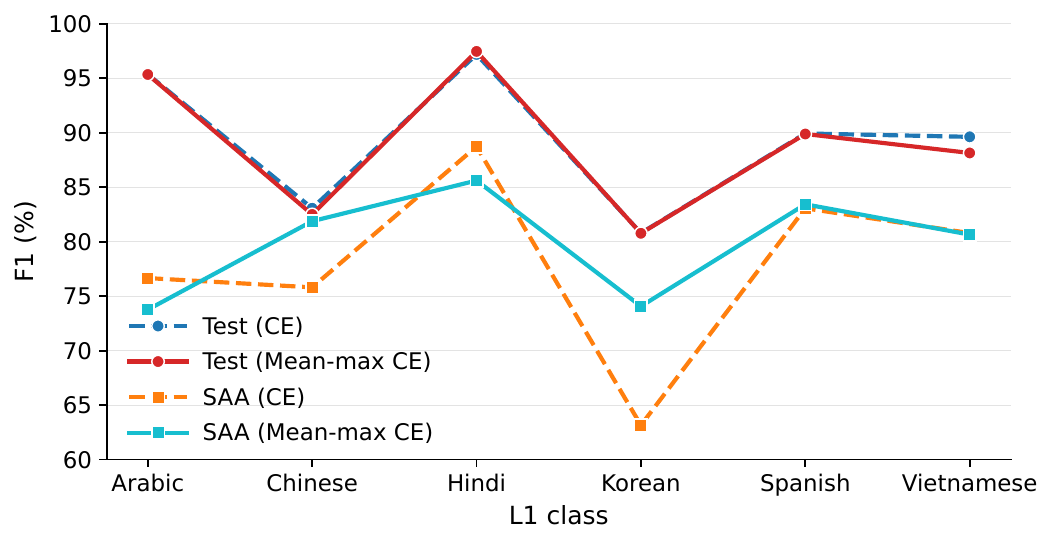}
\caption{\textbf{Class-wise effects of Mean--max CE.}
Per-class F1 on the speaker-disjoint source Test set and SAA under ordinary CE and Mean--max CE ($\alpha=0.25$), averaged over five training seeds.
The four curves represent the two objectives on the two evaluation sets.
On SAA, the range across the six mean class F1 scores decreases from $25.57$ to $11.85$ percentage points.}
\label{fig:classwise-objective}
\end{figure}

Figure~\ref{fig:classwise-objective} and Table~\ref{tab:app-per-class} compare PAID trained with ordinary CE and Mean--max CE ($\alpha=0.25$).
The curves and table report F1 for each L1 class on Test and SAA, averaged over five seeds.

\begin{table}[!htbp]
\centering
\footnotesize
\setlength{\tabcolsep}{2pt}
\caption{\textbf{PAID class F1 under CE and Mean--max CE.}
Mean--max CE uses $\alpha=0.25$.
$N$ is class support and $\Delta$ is Mean--max CE minus CE in percentage points.}
\label{tab:app-per-class}
\begin{tabularx}{\linewidth}{
@{}l r
>{\centering\arraybackslash}X
>{\centering\arraybackslash}X
r@{}}
\toprule
L1 & $N$ & CE & Mean--max CE & $\Delta$ \\
\midrule
\multicolumn{5}{@{}l}{\emph{Test}} \\
Arabic
& 1,951 & $95.36\pm0.25$ & $95.34\pm0.20$ & $-0.02$ \\
Chinese
& 2,389 & $83.07\pm0.61$ & $82.50\pm0.67$ & $-0.57$ \\
Hindi
& 1,296 & $97.18\pm0.18$ & $97.46\pm0.10$ & $+0.27$ \\
Korean
& 2,414 & $80.79\pm1.15$ & $80.77\pm1.18$ & $-0.02$ \\
Spanish
& 1,764 & $89.93\pm0.44$ & $89.88\pm0.79$ & $-0.05$ \\
Vietnamese
& 1,283 & $89.62\pm0.30$ & $88.14\pm0.80$ & $-1.48$ \\
\midrule
\multicolumn{5}{@{}l}{\emph{SAA}} \\
Arabic
& 101 & $76.66\pm4.58$ & $73.76\pm4.92$ & $-2.90$ \\
Chinese
& 65 & $75.81\pm3.17$ & $81.88\pm2.45$ & $+6.07$ \\
Hindi
& 18 & $88.73\pm4.88$ & $85.61\pm8.49$ & $-3.12$ \\
Korean
& 52 & $63.16\pm5.73$ & $74.07\pm4.91$ & $+10.91$ \\
Spanish
& 162 & $83.06\pm0.59$ & $83.43\pm1.56$ & $+0.37$ \\
Vietnamese
& 22 & $80.79\pm6.13$ & $80.65\pm6.74$ & $-0.14$ \\
\bottomrule
\end{tabularx}
\end{table}

On SAA, Mean--max CE increases Korean and Chinese mean F1 by $10.91$ and $6.07$ percentage points, respectively, while Arabic and Hindi decrease by $2.90$ and $3.12$ points.
The range across the six mean class F1 scores decreases from $25.57$ to $11.85$ points on SAA but remains similar on Test ($16.39$ versus $16.69$ points).
The class-wise range is computed after averaging per-class F1 across seeds, unlike mean Worst-F1, which averages the minimum class F1 obtained within each seed.

\section{Implementation and Reporting Details}
\label{app:implementation}

\subsection{Acoustic Features and Phoneme-Label Composition}

Audio is resampled to 16 kHz and truncated to the first 30 seconds.
The frozen Whisper-small encoder supplies 768-dimensional contextual representations at approximately 20-ms intervals.
MFA boundaries are converted to Whisper frame indices and restricted to the processed, non-padding audio region.
The learned acoustic mapping uses Linear--LayerNorm--ReLU--Dropout and outputs 192-dimensional vectors.
The phoneme-label vocabulary contains 41 entries, including padding and unknown symbols.
PAID uses a trainable 64-dimensional phoneme-label embedding, producing a 256-dimensional classifier input after the acoustic and label components are concatenated; the acoustic-only classifier input has 192 dimensions.
% TODO-VERIFY: Exact Whisper output layer, MFA model/dictionary versions,
% time-to-frame rounding and empty-span handling are not documented in the
% supplied manuscript; fill in from configuration and alignment artifacts.

\paragraph{Phoneme-label composition.}
\label{app:phone-composition}
Let $\overline{\mathbf r}$ denote the average transformed acoustic-unit representation, and let $q_p$ be the proportion of retained spans labeled with phoneme $p$.
The mean label embedding is $\sum_p q_p E(p)$.
With dropout disabled, the PAID logits can be decomposed as
\begin{equation}
\mathbf{o}=W_a\overline{\mathbf r}+W_p\sum_p q_pE(p)+\mathbf b,
\label{eq:phone-composition-logits}
\end{equation}
where $W_a$ and $W_p$ are the classifier-weight blocks corresponding to the acoustic and phoneme-label inputs, respectively.
This decomposition describes the contribution of phoneme-label composition to the classifier output.

\subsection{Optimization and Replication}

PAID and the matched Whisper representation controls use AdamW with learning rate $2\times10^{-4}$, weight decay $10^{-4}$, batch size 256, dropout 0.1, and ten epochs.
The five training seeds are 1111, 2222, 3333, 4444, and 5555.
TF--IDF controls use seed 1337 for their single reported run; the phoneme-label-only classifier is trained with five seeds under ordinary CE.
Balanced-sampling controls use inverse class, corpus, or corpus--class frequencies and sample with replacement.

\subsection{Baseline Checkpoints and Reporting Protocols}
\label{app:baseline-implementation}

Whisper-Utt and WavLM-Utt use utterance-level mean pooling followed by a multilayer perceptron (MLP) or logistic-regression classifier.
WavLM-FrameAttn uses a two-layer attention-scoring MLP, ECAPA-Spk uses an embedding followed by logistic regression, and PAID-WavLM replaces Whisper acoustic features with frozen WavLM Base+ features.
The matched Whisper-Global control excludes padding frames when computing the utterance mean.
The original records do not establish the padding-mask rule for the separately trained Whisper-Utt baseline.
Table~\ref{tab:app-checkpoints} lists the pretrained checkpoints.

\begin{table}[!htbp]
\centering
\footnotesize
\setlength{\tabcolsep}{3pt}
\caption{\textbf{Pretrained checkpoints used in the experiments.}}
\label{tab:app-checkpoints}
\begin{tabularx}{\linewidth}{
@{}>{\raggedright\arraybackslash}p{0.28\linewidth}
>{\raggedright\arraybackslash}X@{}}
\toprule
Model & Checkpoint \\
\midrule
Whisper
& \nolinkurl{openai/whisper-small} \\

WavLM
& \nolinkurl{microsoft/wavlm-base-plus} \\

ECAPA-Spk
& \nolinkurl{speechbrain/spkrec-ecapa-voxceleb} \\

CommonAccent-ECAPA
& \nolinkurl{Jzuluaga/accent-id-commonaccent_ecapa} \\

CommonAccent-XLSR
& \nolinkurl{Jzuluaga/accent-id-commonaccent_xlsr-en-english} \\

Voxlect-small
& \nolinkurl{tiantiaf/voxlect-english-}\allowbreak\nolinkurl{dialect-whisper-small} \\

Vox-Profile
& \nolinkurl{tiantiaf/}\allowbreak\nolinkurl{whisper-large-v3-}\allowbreak\nolinkurl{narrow-accent} \\
\bottomrule
\end{tabularx}
\end{table}

Voxlect-small and Vox-Profile supply frozen 256-dimensional features to an MLP with 192 hidden units and six output classes.
Both MLP classifiers use ordinary CE, dropout 0.1, and the five-seed training schedule, with scores reported after epoch 10.
The upstream training for these two checkpoints includes L2-ARCTIC and ESLTTS.
Table~\ref{tab:app-protocols} summarizes the reporting conventions used for the different system groups.

\begin{table}[t]
\centering
\small
\setlength{\tabcolsep}{2pt}
\caption{\textbf{Reporting protocols.}}
\label{tab:app-protocols}
\begin{tabularx}{\linewidth}{
@{}>{\raggedright\arraybackslash}X
>{\raggedright\arraybackslash}p{0.24\linewidth}
>{\raggedright\arraybackslash}p{0.29\linewidth}@{}}
\toprule
Experiment & Replication & Reporting \\
\midrule
PAID and acoustic-unit controls
& Five seeds
& Final epoch 10 \\
Source controls
& Five seeds
& Final epoch 10 \\
Voxlect / Vox-Profile
& Five seeds
& Final epoch 10 \\
Other pretrained baselines
& One or five runs
& Development-selected \\
TF--IDF controls
& One run (seed 1337)
& Development-selected regularization \\
\bottomrule
\end{tabularx}
\end{table}

\subsection{Dataset Composition}
\label{app:dataset-coverage}

The six-L1 source dataset is constructed from L2-ARCTIC and ESLTTS by harmonizing L1 labels, organizing corpus-qualified speaker identities, and defining speaker-disjoint Train, Development, and Test partitions.
Each of the six labels appears in both source corpora in every partition.
Table~\ref{tab:app-dataset-composition} gives the speaker counts by source and the retained utterance counts by L1 and partition.

\begin{table}[!htbp]
\centering
\small
\setlength{\tabcolsep}{2pt}
\caption{\textbf{Six-L1 dataset composition.}
Upper block: Train/Dev/Test speaker counts.
Lower block: utterances retained after alignment and feature preprocessing.}
\label{tab:app-dataset-composition}

\begin{tabular*}{\linewidth}{@{\extracolsep{\fill}}lrrr@{}}
\toprule
L1 & L2-ARCTIC & ESLTTS & Unified \\
\midrule
Arabic & 2/1/1 & 2/1/1 & 4/2/2 \\
Chinese & 2/1/1 & 10/2/3 & 12/3/4 \\
Hindi & 2/1/1 & 6/1/1 & 8/2/2 \\
Korean & 2/1/1 & 8/2/2 & 10/3/3 \\
Spanish & 2/1/1 & 15/3/4 & 17/4/5 \\
Vietnamese & 2/1/1 & 4/1/1 & 6/2/2 \\
\midrule
Total & 12/6/6 & 45/10/12 & 57/16/18 \\
\bottomrule
\end{tabular*}

\medskip

\begin{tabular*}{\linewidth}{@{\extracolsep{\fill}}lrrrr@{}}
\toprule
L1 & Train & Dev & Test & SAA \\
\midrule
Arabic & 4,248 & 1,730 & 1,951 & 101 \\
Chinese & 6,404 & 1,445 & 2,389 & 65 \\
Hindi & 5,549 & 1,878 & 1,296 & 18 \\
Korean & 3,475 & 2,421 & 2,414 & 52 \\
Spanish & 6,486 & 2,540 & 1,764 & 162 \\
Vietnamese & 5,210 & 2,173 & 1,283 & 22 \\
\midrule
Total & 31,372 & 12,187 & 11,097 & 420 \\
\bottomrule
\end{tabular*}
\end{table}

\subsection{Split Provenance and Statistical Aggregation}
\label{app:provenance}

Corpus-qualified speaker identifiers define the source partitions, so no source speaker belongs to multiple Train, Development, or Test splits.
Before alignment and acoustic-feature preprocessing, the source Train, Development, and Test partitions contain 31,390, 12,188, and 11,224 recordings, respectively; 421 SAA recordings are in the raw external set.
The corresponding retained counts are 31,372, 12,187, 11,097, and 420.
The curated source set therefore contains 54,656 retained utterances from 91 speakers, and the external SAA set contains 420 utterances from 420 distinct speakers.
The original \texttt{mandarin} label is mapped to Chinese in the common six-class label space.
% TODO-VERIFY: Add exclusion reasons/counts and the retained-recording manifest
% identifiers only after checking the original preprocessing audit.

Metrics for five-seed configurations are computed for each seed before the mean and sample SD are reported.
If $F_{s,c}$ denotes the F1 for class $c$ under seed $s$, the reported mean Worst-F1 is $S^{-1}\sum_{s=1}^{S}\min_c F_{s,c}$, with $S=5$.
The class-wise F1 range is $\max_c\overline{F}_c-\min_c\overline{F}_c$, where $\overline{F}_c=S^{-1}\sum_s F_{s,c}$.
These two aggregations are calculated in different orders.
Balanced accuracy is defined in Equation~\ref{eq:balanced-accuracy} in the main text.

\end{document}